\documentclass[aip,jap,reprint,amsmath,amssymb,floatfix,groupedaddress]{revtex4-1}
\pdfoutput=1
\usepackage{graphicx}  
\usepackage{color}
\usepackage{hyperref} 
\usepackage{dcolumn}
\usepackage{bm}

\graphicspath{{./figures/}}

\newcommand{\mr}[1]{\mathrm{#1}}

\newcommand{\Si}{\ensuremath{\mathrm{ ^{28} Si}}}

\newcommand{\al}{\textit{et al.}}

\newcommand{\nm}{\ensuremath{\mathrm{nm}}}

\newcommand{\m}{\ensuremath{\mathrm{m}}}

\newcommand{\fs}{\ensuremath{\mathrm{fs}}}	

\newcommand{\ns}{\ensuremath{\mathrm{ns}}}

\newcommand{\GW}{\ensuremath{\mathrm{GW}}}

\newcommand{\K}{\ensuremath{\mathrm{K}}}

\newcommand{\degr}{\ensuremath{\mathrm{^{\circ}}}}

\newcommand{\TCu}{\ensuremath{\mathrm{W \, m^{-1} \, K^{-1}}}}

\newcommand{\RKu}{\ensuremath{\mathrm{K\,m^2/GW}}}

\newcommand{\RK}{\ensuremath{R_\mr{K}}}

\newcommand{\Fig}[1]{Fig.~\ref{fig:#1}}
\newcommand{\FigB}[1]{Figure~\ref{fig:#1}}
\newcommand{\fig}[1]{\ref{fig:#1}}

\newcommand{\Tab}[1]{Table~\ref{tab:#1}}

\newcommand{\Eq}[1]{Eq.~(\ref{eq:#1})}

\newcommand{\Sub}[1]{Sec.~\ref{sub:#1}}

\begin{document}

\title
{Thermal resistance of grain boundaries in silicon nanowires by
nonequilibrium molecular dynamics}

\author{Jan K. Bohrer}
\author{Kevin Schr\"oer}
\email{kevin.schroeer@uni-due.de}
\author{Lothar Brendel}
\author{Dietrich E. Wolf}
\affiliation{Department of Physics and Center for Nanointegration Duisburg-Essen (CENIDE),
Universit\"at Duisburg-Essen, D-47048 Duisburg, Germany}

\date{\today}

\begin{abstract}
  The thermal boundary resistance (Kapitza resistance) of (001) twist
  grain boundaries in silicon nanowires depends on the mismatch angle.
  This dependence is systematically investigated by means of
  nonequilibrium molecular dynamics simulations. Grain boundary systems
  with and without coincidence site lattice are compared. The Kapitza
  resistance increases with twist angle up to $40\degr$. For larger
  angles, it varies only little around $1.56\pm 0.05~\RKu$,
  except for a drop by $30\%$ near the $90\degr$ $\Sigma 1$ grain
  boundary. Finite size effects due to the fixed outer boundary
  conditions of the nanowire are negligible for diameters larger than
  $25~\nm$. 


\end{abstract}


\maketitle

\section{\label{sec:Intro}Introduction}

Due to its thermal and electronic transport properties, it is
a promising approach to use nanostructured silicon as thermoelectric
material.\cite{Schierning2014} The efficiency of a thermoelectric device at temperature
$T$ increases with the material's figure of merit $zT$,
\begin{equation}
	zT = \frac{\alpha^2 \sigma}{\kappa} T \, ,
	\label{eq:}
\end{equation}
where $\alpha$ is the Seebeck coefficient, $\sigma$ the electrical conductivity, and
$\kappa$ the thermal conductivity. As $z$ consists only of material properties,
target-oriented material design is the consequent approach to improve the efficiency of
thermoelectric generators. 

Evidently, it is desirable to find materials with a high Seebeck coefficient, a high
electrical conductivity, and a low thermal conductivity. The approach in this work
is to reduce the phononic contribution to heat conduction without significantly decreasing
the electronic transport. This is possible in crystalline silicon, because $50\%$ of the
thermal transport is carried by phonons with mean free paths of $500~\nm$ and
longer.\cite{Regner2013} Electrons, on the other hand, have effective mean free paths on a
much shorter scale.\cite{Schierning2014} Due to advances in processing, highly doped
polycrystalline silicon with crystallite sizes in the order of $25~\nm$ can be prepared
from nanoparticles.\cite{Schierning2011} For structures on the nano scale, the phonon
scattering is expected to change, making phonon-surface scattering the most important
mechanism.\cite{Schierning2014} The expected significant reduction in thermal conductivity
was achieved by producing silicon materials consisting of nanosized crystalline
grains.\cite{Schierning2011}

While the thermal conductance, and thus the average influence of grain boundaries (GBs) in
polycrystalline silicon films, was measured experimentally,\cite{Uma2001} the investigation
of actual individual grain boundaries is challenging. Therefore, molecular dynamics (MD)
simulations can be of great help to elucidate transport processes and estimate the
magnitude and behavior of thermal material properties.\cite{Maiti1997, Schelling2002,
Schelling2004, Kimmer2007, Watanabe2007,daCruz2011}

Mainly, we will investigate the dependence of thermal resistance at twist grain
boundaries on the twist angle. Based on results of former MD simulations for silicon and
diamond grain boundaries,\cite{Schelling2004, Watanabe2007, Ju2013} we expect interfacial
resistances in the order of a few $\RKu$ and an increase with twist angle up to at least
$40\degr$. Using systems of finite size with fixed boundary conditions, we also aim at
simulating the thermal processes between two actual grains of a given size more
accurately as for periodic boundary conditions. On the other hand, we will need to estimate
inadequacies due to the artificial fixed atomistic walls.

\section{\label{sec:Meth}Methods}
Thermal transport in silicon is simulated by nonequilibrium molecular dynamics (NEMD)
using the Stillinger-Weber potential with original parameters,\cite{Stillinger1985} 
which is well characterized and was shown to produce realistic results in previous cases.
\cite{Schelling2004,Wang2009,daCruz2011,Ju2013} For
our simulations, we used the well established program LAMMPS.\cite{Plimpton1995}

\begin{figure*}[htbp]
\begin{center}
  \includegraphics[]{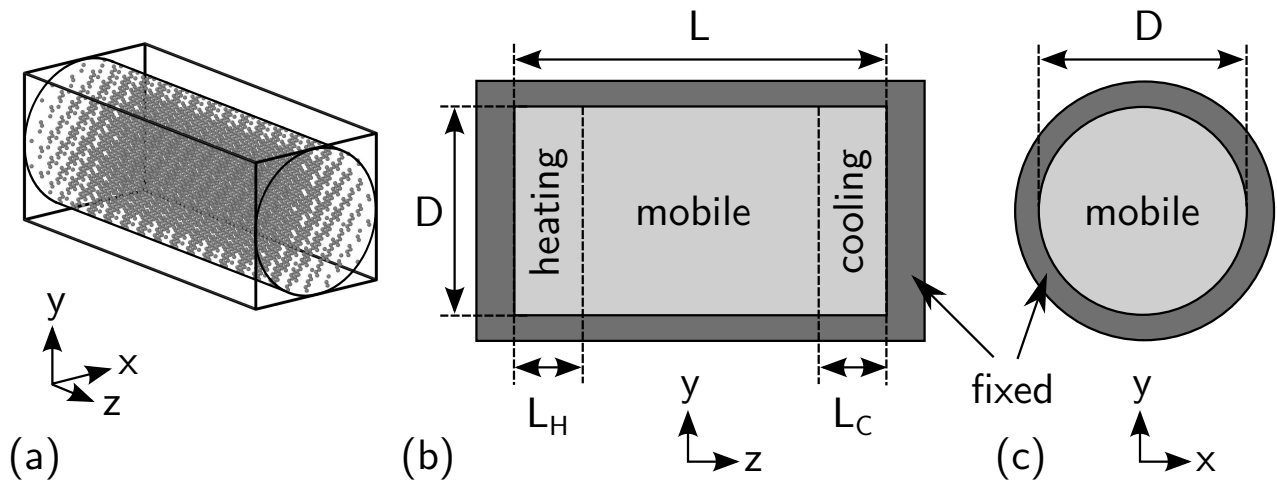}
  \caption{\label{fig:Cyl_sketch}Simulated system with fixed atomistic
  boundary conditions: (a) The simulated mono-crystalline nanowire of cylindrical shape
  contains $\Si$ isotopes, which are initially placed at their equilibrium positions at
  $T=0~\K$. (b, c) The Si-wire is completely enclosed in a cylindrical shell of
  fixed Si-atoms.
  The dimensions of the mobile part are diameter $D$ and length $L$, which includes the
  heating and cooling layers (lengths $L_{H,C}$), where kinetic energy is injected and
  withdrawn respectively. The z-axis is parallel to the [001]-direction of the crystal
  lattice.}
\end{center}
\end{figure*}

To obtain rotationial symmetry, we choose cylindrical geometry for the investigated
nanowires, as shown in \Fig{Cyl_sketch}. The fixed atomistic walls have a thickness of
$1.2$~lattice constants for the cylindrical shell and one lattice constant at both ends.
Both the mobile part and the fixed walls consist of isotopically pure $\Si$. Particle-wall
interactions are described by the same potential as particle-particle interactions.

The following method is used to establish a constant one dimensional heat flux $j$
through the system (along $z$):
A slab of the nanowire (thickness $L_\mr{H}$) at one end is heated by
adding a certain amount of kinetic energy $\Delta K$ per time step $\Delta t$
to the atoms by rescaling their velocity vectors accordingly. The slab at
the other end (thickness $L_\mr{C} = L_\mr{H}$) is cooled by the same method, extracting
the same amount of kinetic energy per time step and thus maintaining energy conservation of the
whole system. The heating and cooling regions are scaled with the system length $L$
having always a thickness of $L/22$.

In stationary state, the heat flux density is homogeneous and amounts to
\begin{equation}
	j = \frac{\Delta K}{\Delta t\,A} \, ,
	\label{eq:j_def}
\end{equation}
where $A$ is the cross sectional area of the cylindrical nanowire. 

The local instantaneous temperature $T$ is defined by dividing the wire into slabs (with thickness 
of one lattice constant $a_0$) and respectively averaging the kinetic energy
$K_\mr{int}$ of the contained inner degrees of freedom $N_\mr{int}$ at a given time step,
\begin{equation}
	T := \frac{2\,K_\mr{int}}{N_\mr{int}\,k_\mr{B}} \, , 
	\label{eq:temp_def}
\end{equation} 
where $k_\mr{B}$ is the Boltzmann constant. This instantaneous temperature is then averaged over
a time interval to produce a temperature profile $T(z)$, which, in the stationary state, does
not depend on time except for statistical fluctuations.

Initially, the atoms are placed at the $0~\K$ equilibrium positions of the silicon lattice
and are given random velocities corresponding to a Maxwell-Boltzmann distribution with
a temperature of $300~\K$. Thereafter, the MD simulation with a time step of $1~\fs$ 
consists of two phases: First, the system is equilibrated for $3~\ns$ at $300~\K$ using the
Nos\'e-Hoover-thermostat\cite{Nose1984} with constant volume $V$ and constant number of
particles $N$ ($NVT$-ensemble). For large systems ($L > 170~a_0$ or $A \geq 200~a_0^2$)
this equilibration time is extended to $4~\ns$. Second, the heat flux is established and
maintained for $6~\ns$.
In this second phase, the system is given a time of $3~\ns$ to reach the stationary state and
the last $3~\ns$ are used to measure and average the temperature profiles. The heat flux
density is chosen to be the same for all investigated systems ($j = 21.73~\GW/\m^2$).

\subsection{\label{sub:methods_kappa}Measuring thermal conductivity}

At first, we investigate the thermal properties of crystalline nanowires devoid of any 
grain boundary (GB). Knowing the heat flux and the temperature profile $T(z)$ from the
simulation, the thermal conductivity $\kappa$ is determined from Fourier's law (in one
dimension):
\begin{equation}
	j = -\kappa \frac{dT}{dz} \, .
	\label{eq:Fourier}
\end{equation}

\begin{figure}[htbp]
\begin{center}
  \includegraphics[]{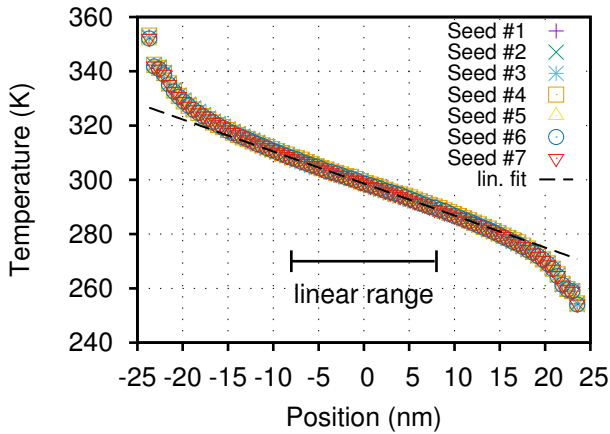}
  \caption{\label{fig:kappa_method}Temperature profiles averaged over 3~ns for a constant
  heat flux through a system of $L=47.8~\nm$ and $D=6.13~\nm$ without grain boundary
  in stationary state. For the same geometrical system, different initial velocity
  distributions constituting the same initial temperature were generated by using
  different seeds in a random number generator. Simulation results for seven different
  initial velocities (seeds) show minor variations in the temperature profiles. The
  thermal conductivity is calculated from the slope of the averaged profile in the linear
  range.}
\end{center}
\end{figure}

If Fourier's law is valid and $j$ is constant, we expect a linear temperature profile,
provided $\kappa$ does not vary too much spatially. \FigB{kappa_method} shows that
linearity is given in the middle of the nanowire. At the ends, where the
heating and cooling takes place, the system is not in local equilibrium and Fourier's law
does not apply. In the heat source and sink, the system is externally perturbed by
rescaling the velocities and can not reach local equilibrium, which also influences the
adjacent regions. The nonlinear behavior was reported by other
authors,\cite{Schelling2002,daCruz2011} and accounted for by the modified phonon
scattering at the heating and cooling regions.

Therefore, the profiles are fitted in the linear range and $\kappa$ is calculated
from \Eq{Fourier}. To proof consistency, we simulate the same system several times, only
changing the seed of the random number generator that creates the initial velocity
distribution. Although the different profiles in \Fig{kappa_method} do not vary much in
the graphical representation, numerical evaluation shows that the thermal conductivities
vary about 1\%.

To account for these statistical variations, we average the temperature profiles of the
different seeds, which leads to a mean profile with error bars. The slope in the linear range and its uncertainty are then
determined from the mean profile and its error bars.

\subsection{\label{sub:methods_RK} Grain boundaries and Kapitza resistance}

\begin{figure}[htbp]
\begin{center}
  \includegraphics[]{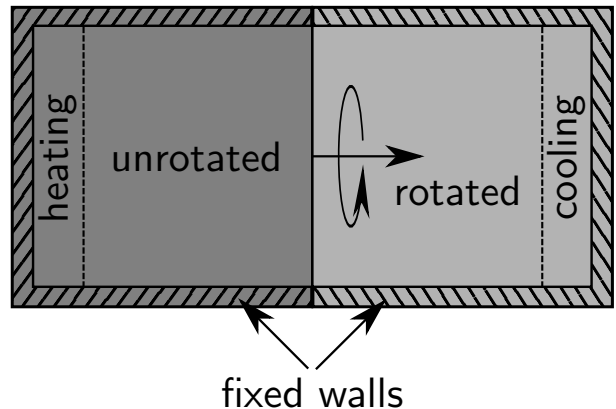}
  \caption{\label{fig:GB_sketch}One half of the cylinder, i.e., all atoms with positions
  $z > 0$, are rotated by an angle $\vartheta$ around an [001]-axis situated in the center
  of the cylinder cross section. The (001)-rotation plane lies in the x-y-plane and
  defines the position of the grain boundary. The rotation includes the affected atoms of
  the fixed walls.}
\end{center}
\end{figure}

To implement a twist grain boundary of a certain angle, one half of the cylindrical
mono-crystalline nanowire is rotated with respect to the other, including the fixed walls,
by a twist angle $\vartheta$, as can be seen in \Fig{GB_sketch}. The rotation plane (x-y)
is an (001)-plane of the diamond lattice. This prescribes the initial atom configuration. We
simulate GB systems with and without a coincidence site lattice (CSL).\cite{Santoro1973} The
former are called $\Sigma$ grain boundaries. $\Sigma$ corresponds to the size of the primitive unit
cell of the CSL relative to the original primitive unit cell of silicon.

We choose fixed boundary conditions, because we want to be able to investigate finite size
effects and arbitrary grain boundary twist angles. For periodic boundary conditions, only
some specific angles, which create a coincidence site lattice, would be
possible to simulate.\cite{Schelling2004} Furthermore,
rotation symmetry is necessary for fixed boundary conditions to obtain equivalent regions on
both sides of the grain boundary planes. This, in the end, is the reason for using systems
with cylindrical geometry.

\begin{figure}[htbp]
\begin{center}
  \includegraphics[]{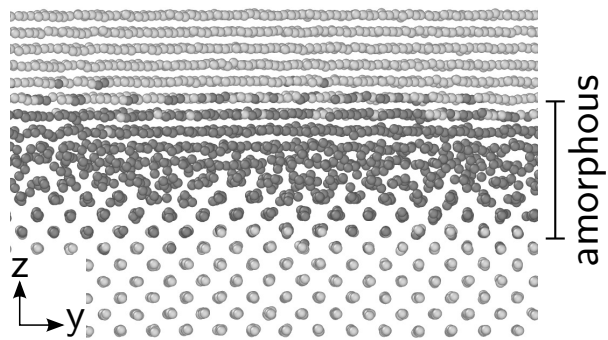}
  \caption{\label{fig:amorph}After thermostating
  the $36.87\degr$ $\Sigma 5$ grain boundary system to 300~K, the
  atomic structure is examined by using common neighbor analysis. The atoms in light
  gray have diamond lattice structure, while the dark gray atoms have no crystalline
  order and constitute the amorphous region of the grain boundary.}
\end{center}
\end{figure}

During the thermostat phase, an amorphous region emerges at the grain boundary. Common
neighbor analysis, as proposed by Honeycutt and Andersen,\cite{Honeycutt1987} shows, which
atoms are considered to be in an crystalline environment of the diamond lattice and which
have no crystalline order (see \Fig{amorph}). The thickness of the amorphous region is
estimated and lies between one and three lattice constants, depending on the twist angle
and system dimensions.

\begin{figure}[htbp]
\begin{center}
  \includegraphics[]{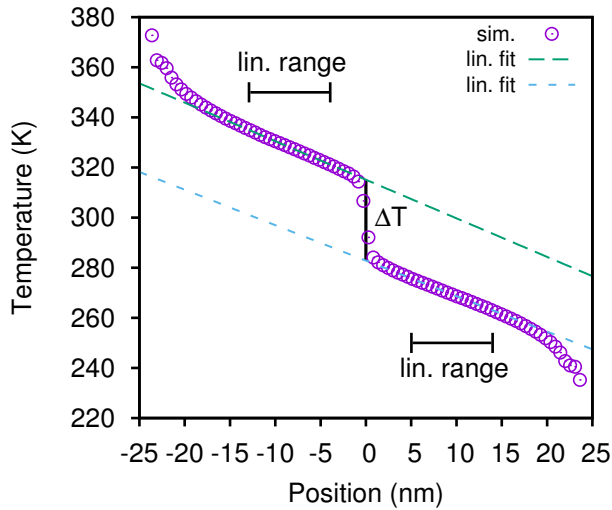}
  \caption{\label{fig:RK_method}Measured temperature profile (sim.) averaged over 3~ns for
  a $36.87\degr$ $\Sigma 5$ grain boundary system of $L=47.8~\nm$ and $D=6.13~\nm$. The
  temperature gap $\Delta T$ is determined by extrapolating the two linear ranges and
  taking the difference of the linear fits at the grain boundary interface.}
\end{center}
\end{figure}

The grain boundaries constitute a resistance to thermal transport, called Kapitza
resistance $\RK$. In the presence of a heat flux, it leads to a temperature discontinuity
$\Delta T$ at the interface,
\begin{equation}
	\Delta T = \RK \, j \, ,
	\label{eq:RK_def}
\end{equation}
which is visible in the simulated profiles (see \Fig{RK_method}).

The temperature jump is determined by extrapolating the linear ranges on both sides of the
grain boundary and taking the difference of the linear fits at the interface (see
\Fig{RK_method}). Knowing the constant heat flux density, the Kapitza resistance is then
calculated using \Eq{RK_def}.

\begin{figure}[htbp]
\begin{center}
  \includegraphics[]{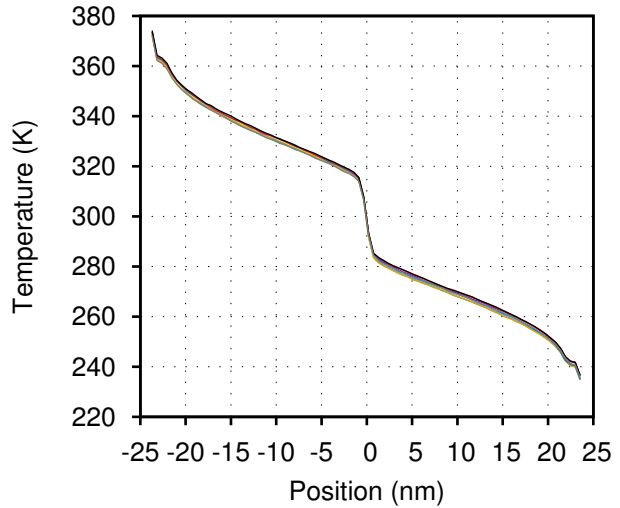}
  \caption{\label{fig:RK_seeds}Temperature profiles for a $36.87\degr$ $\Sigma 5$ grain
  boundary system of $L=47.8~\nm$ and $D=6.13~\nm$ for nine separate simulation runs with
  different initial velocity distributions (seeds).}
\end{center}
\end{figure}

Again, the consistency of the results is checked by comparing simulation runs with
different initial velocity distributions. Although the graphical representation shows 
no major visible differences of the temperature profiles (see \Fig{RK_seeds}), the numerical evaluation of the temperature gap leads to
significant deviations between the runs. Therefore the profiles are averaged, as for the
GB-free systems (cf. \Sub{methods_kappa}), to create a mean profile with statistical error
bars for each data point. The Kapitza resistance and its uncertainty are determined from
the linear fits to this mean profile.

\section{\label{sec:Res}Results and discussion}

In this section, we analyze finite size effects on the thermal properties of systems with
and without grain boundary to discuss the influence of the chosen geometry and fixed
boundary conditions. We then present our results of the Kapitza resistance measurements
for GB systems with 33 different twist angles.
 
\subsection{\label{sub:FS_kappa}Finite size effects on the thermal conductivity}

\begin{figure}[htbp]
\begin{center}
  \includegraphics[]{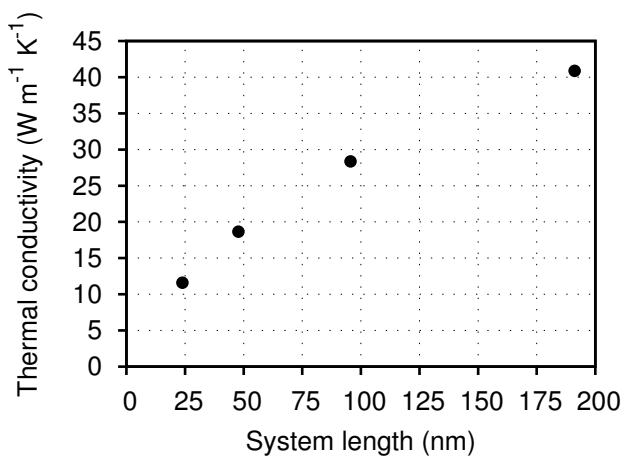}
  \caption{\label{fig:kappa_L}Variation of thermal conductivity with system length for
  a system without grain boundary of constant diameter of $D=6.13~\nm$ at $T=300~\K$.
  Error bars are smaller than $1.8~\%$ and not visible in this representation.}
\end{center}
\end{figure}

The mono-crystalline, cylindrical system is simulated for four different wire lengths
from $23.9~\nm$ to $191~\nm$ with a constant diameter of $6.13~\nm$. \FigB{kappa_L} shows
that the thermal conductivity increases with length and (for our system) lies between $11.6~\TCu$ and
$40.9~\TCu$. Schelling \textit{et al.} reported a similar behavior (increase of $\kappa$
with system length) for periodic boundary conditions due to the reduction of the phonon
mean free path.

\begin{figure}[htbp]
\begin{center}
  \includegraphics[]{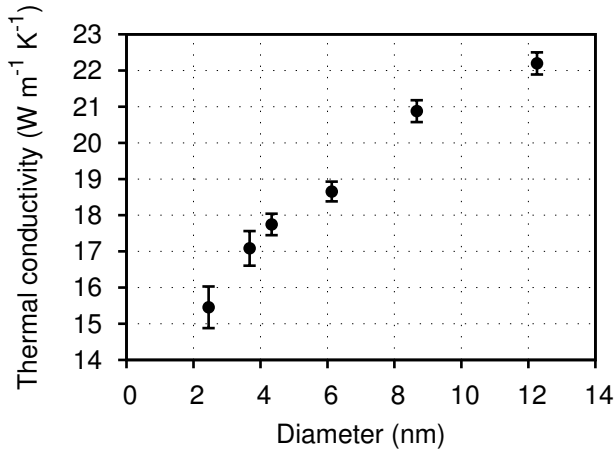}
  \caption{\label{fig:kappa_D}Variation of thermal conductivity with diameter for
  a system without grain boundary of length $L = 47.8~\nm$ at $T=300~\K$.}
\end{center}
\end{figure}

Likewise, we vary the diameter six times between $2.45~\nm$ and $12.26~\nm$ for a constant
length of $47.8~\nm$ to investigate the influence of the cross section. As can be seen in
\Fig{kappa_D}, the thermal conductivity also increases with diameter and attains values
between $15.5~\TCu$ and $22.2~\TCu$. However, the system length's influence on the heat
conduction is much stronger than the diameter's.

It is well known from experiment and simulation that heat conduction is
reduced in nano-sized and nano-crystalline media.\cite{Schelling2002,Li2003,daCruz2011}
By comparing our results to experimental values of $\kappa \approx 250~\TCu$ for
bulk $\Si$ \cite{Capinski1997}, we can confirm a reduction of the thermal
conductivity in nanowires by one to two orders of magnitude.

Da Cruz \textit{et al.}\cite{daCruz2011} used NEMD to study the thermal transport in nanowires
of square cross section for different potentials and boundary conditions. They reported
similar finite size effects, i.e., increase of the thermal conductivity with length and cross
sectional area $A$. For fixed atomistic boundary conditions and and the Stillinger-Weber potential, they
measured $\kappa \approx 15.5~\TCu$ for a length of $27.2~\nm$ and cross section of
$4.34~\nm \times 4.34~\nm = 18.8~\nm^2$.

For a further investigation and comparison, we simulate a nanowire with square cross section
of $A = 29.5~\nm^2$ and $L = 23.9~\nm$, leading to $\kappa = 15.2~\TCu$. Extrapolating to
the dimensions used by da Cruz \textit{et al.} (stated above), we obtain $\kappa =
15.9~\TCu$, which is in good agreement with their result of $15.5~\TCu$.

Comparing the cuboidal setup with our cylindrical system of same length and cross
sectional area, i.e. $L = 23.9~\nm$ and $D = 6.13~\nm$, we note a reduction in the
thermal conductivity to $11.6~\TCu$. Thus, by solely changing the shape of the cross
section from square to circular, the heat conduction is reduced by $\sim 25\%$. We
attribute this to inadequacies of the MD simulation, especially in combination with fixed
boundary conditions: 

For the square cross section, the mobile part of the system consists of entire cubic unit
cells. For the cylindrical geometry the unit cells near the walls will be cut into regions
of mobile and fixed atoms. We propose that the reduction in thermal conductivity is a
consequence of the less regular and less periodic particle-wall bonds, which introduce
additional phonon scattering. Additionally, da Cruz \al\cite{daCruz2011} observed that,
for simulated nanowires, the thermal conductivity is about twice as high for fixed
boundary conditions as for free boundary conditions in the transversal (x-y) directions.

Due to the challenge of characterizing nano scale systems experimentally and the
imperfect material properties, comparison with real systems is difficult. Li
\al\cite{Li2003} investigated isotopically natural, intrinsic, crystalline Si nanowires
with diameters of $22~\nm$ and a length of several microns, measuring thermal conductivities
of about $7~\TCu$ at $300~\K$. Moreover, the results of Boukai \al\cite{Boukai2008}
indicated experimental values of about $0.8~\TCu$ for highly doped Si nanowires with cross
sectional areas of $10~\nm \times 20~\nm$, and of roughly $3.5~\TCu$ for $A = 20~\nm
\times 20~\nm$. Here, the wires were also several microns long.

We compare those values with $\kappa = 22.2~\TCu$ for our largest simulated diameter of
$12.3~\nm$ (and $L = 47.8~\nm$). Based on the preceding considerations, we conclude an
overestimation of the heat conduction in our simulated systems compared to natural and doped
Si nanowires. Reasons are the artificial fixed boundary conditions, the isotopic purity,
the perfection of the crystal lattice and material surface without impurities or
oxidation, and, most importantly for small systems, the disregard of quantum mechanical
effects, which will lead to significant variations, even for a temperature of
$300~\K$.\cite{Wang2009}

Nevertheless, NEMD can give important information for thermal processes under perfect
conditions and indications of the thermal properties. Moreover, it is most useful to
investigate the influence of parameter variations by tendency, like scaling dimensions,
or, in our case, varying the grain boundary twist angle (see \Sub{RK_Th}).

\subsection{\label{sub:FS_RK}Finite size effects on the Kapitza resistance}

\begin{figure}[htbp]
\begin{center}
  \includegraphics[]{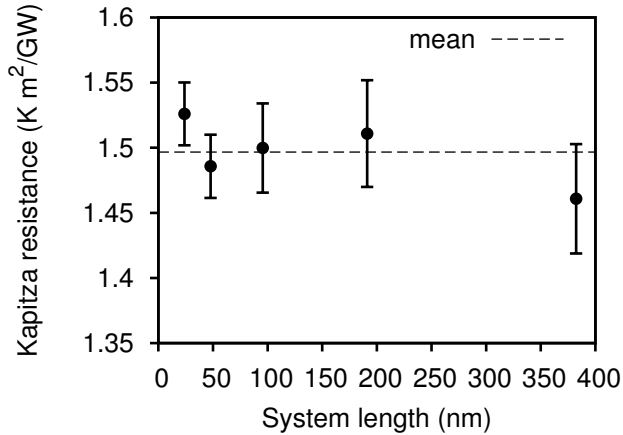}
  \caption{\label{fig:RK_L}Variation of the Kapitza
  resistance with system length for a $36.87\degr$ $\Sigma 5$ grain boundary system with
  a constant diameter of $D=6.13~\nm$ at $T=300~\K$.}
\end{center}
\end{figure}

Next, we present our results for the simulated grain boundary systems, investigating the
influence of length, diameter, and, in \Sub{RK_Th}, twist angle variations on the thermal
boundary resistance. The $36.87\degr$ $\Sigma 5$ GB system with $D = 6.13~\nm$ is
simulated for five different lengths from $23.9~\nm$ to $382~\nm$, leading to a Kapitza
resistance in the range from $1.46~\RKu$ to $1.53~\RKu$ (see \Fig{RK_L}). Although the
shortest and longest system have the highest and lowest resistance, respectively, we can
not conclude a systematic trend in the light of the given uncertainties. In fact, except
for the highest value, the deviation from the mean is covered by the error bars and the
highest deviation amounts to $2.4\%$ (see \Fig{RK_L}). While we can not exclude a possible
influence of the system length on the Kapitza resistance, the results state at any rate
that it is of minor importance for the considered dimensions. Schelling
\al\cite{Schelling2004} used NEMD with periodic boundary conditions to study heat
conduction in Si $\Sigma$ grain boundaries. They also report very little effect of length
variations on the Kapitza resistance for the $43.6\degr$ (001) GB system. 

\begin{figure}[htbp]
\begin{center}
  \includegraphics[]{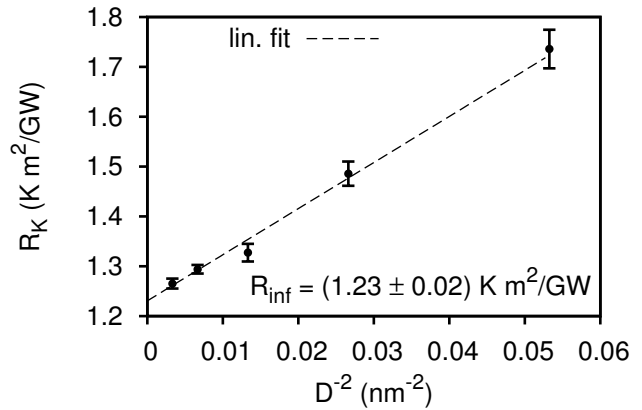}
  \caption{\label{fig:RK_D_inv}Variation of the
  Kapitza resistance $\RK$ with diameter $D$ for a $36.87\degr$ $\Sigma 5$ grain
  boundary system with a constant length of $L = 47.8~\nm$ at $T=300~\K$. The Kapitza
  resistance shows a linear behavior with inverse squared diameter. The value for infinite
  cross section may be estimated by extrapolation.}
\end{center}
\end{figure}

A Variation of the diameter, on the other hand, leads to a strong and systematic behavior of the
Kapitza resistance. For the same twist angle of $36.87\degr$ and a constant length of
$47.8~\nm$, we use five different diameters from $4.33~\nm$ to $17.3~\nm$. As can be seen
in \Fig{RK_D_inv}, the Kapitza resistance shows a decreasing and converging trend with
increasing cross sectional area. Plotting $\RK$ versus inverse squared diameter leads to
data that is in good agreement with a linear fit. Although we did not establish a
theory to explain this behavior, we justify the approach
\begin{equation}
	\RK = R_\mr{inf} + \frac{\gamma}{A} \, ,
	\label{eq:RK_D}
\end{equation} 
with $A = \pi\,D^2/4$, and a coefficient $\gamma = 7.26~\nm^2 \, \m^2 \, \K/\GW$, by the high
quality of the fit.

Furthermore, we extrapolate to infinite cross sectional area ($R_\mr{inf} = 1.23 \pm
0.02~\RKu$) to investigate the influence of the fixed boundary conditions. Our
argument is that for $A\to\infty$, all types of lateral boundary conditions (free,
fixed, and periodic in x-y-directions) should converge to the same results. From the
extrapolation, we can estimate how big the modification due to the fixed boundaries can
be at the most. \Tab{RK_D} shows that the deviation of the Kapitza resistance from
$R_\mr{inf}$ is $5\%$ and less for diameters of $D > 12~\nm$. We propose that for larger
diameters, the differences in $\RK$ caused by different lateral boundary conditions should
become sufficiently small.

\begin{table}[htbp]
\caption{\label{tab:RK_D}Kapitza resistance of the simulated $36.87\degr$ $\Sigma 5$ GB
systems with a constant length of $L = 47.8~\nm$ for five different cross sections.
We include the deviation from the extrapolated resistance $R_\mr{inf} = 1.23~\RKu$ for
infinite cross section.}
\begin{ruledtabular}
\begin{tabular}{ddcc}
\multicolumn{1}{c}{$D~(\nm)$} & \multicolumn{1}{c}{$A~(\nm^2)$} & $\RK~(\RKu)$ &
$(\RK-R_\mr{inf})/\RK$ \\
\hline
4.33 & 14.7 & $1.74 \pm 0.04$ & 0.29 \\ 
6.13 & 29.5 & $1.49 \pm 0.02$ & 0.17 \\ 
8.67 & 59.0 & $1.33 \pm 0.02$ & 0.08 \\ 
12.3 & 118 & $1.29 \pm 0.01$ & 0.05 \\ 
17.3 & 236 & $1.27 \pm 0.01$ & 0.03 \\ 
\end{tabular}
\end{ruledtabular}
\end{table}

However, due to the high computational effort for large systems, we use $D = 6.13~\nm$ and $L =
47.8~\nm$ to investigate the twist angle's influence in \Sub{RK_Th}. If necessary, we are
then able to extrapolate to infinite cross section by reducing the boundary resistance by
about $17\%$. The finite size analysis was done only for the presented $36.87\degr$ GB.
Nevertheless, we expect similar behavior for other twist angles, and use the same scaling
law as a first approach.

\subsection{\label{sub:RK_Th}Dependence of Kapitza resistance on the grain boundary angle}

\begin{figure}[htbp]
\begin{center}
  \includegraphics[width=1.01\linewidth]{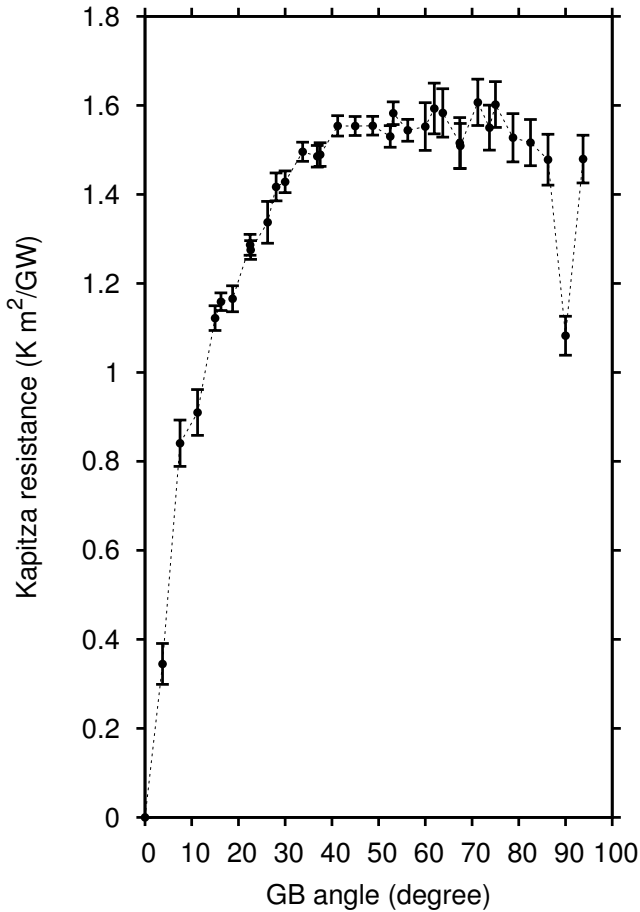}
  \caption{\label{fig:RK_Th}Dependence of the Kapitza resistance on the grain boundary twist
  angle for a system length of $L = 47.8~\nm$ and diameter of $D = 6.13~\nm$ at
  $T=300~\K$.}
\end{center}
\end{figure}

We investigate 25 twist angles from $3.75\degr$ to $93.75\degr$ in equal steps
of $3.75\degr$. Due to the twofold symmetry of the diamond lattice, angles of
$90\degr\pm\phi$ have equivalent configurations and should lead to the same results.
This is verified for our simulations by comparing the $86.25\degr$ and $93.75\degr$ angle
systems, which indeed yield the same Kapitza resistance of $1.48~\TCu$ (see \Fig{RK_Th}).

Additionally, we simulate eight GB systems with coincidence site lattices. Here, the five
smallest $\Sigma$ values are chosen for investigation. Thus, including the
$90\degr$ $\Sigma 1$ GB, we study nine $\Sigma$ grain boundaries (see \Tab{Sigma}) and a
total of 33 GB systems. The determined values of the thermal boundary resistance for all
angles are summarized in \Tab{RK_Th}.

\begin{table}[b]
\caption{\label{tab:Sigma} The nine investigated $\Sigma$ grain boundaries. There are two
twist angles ($\vartheta_{1,2}$) for every $\Sigma$ value, symmetric about $45\degr$.
As a trivial case, the $0\degr$ mono-crystalline system has no grain boundary and is
listed in parenthesis for reasons of clarity.}
\begin{ruledtabular}
\begin{tabular}{rrr}
$\Sigma$ & $\vartheta_1~(\degr)$  & $\vartheta_2~(\degr)$ \\
\hline
25 & 16.26 & 73.74 \\ 
13 & 22.62 & 67.38 \\ 
17 & 28.07 & 61.93 \\ 
5 & 36.87 & 53.13 \\
1 & (0.00) & 90.00 \\
\end{tabular}
\end{ruledtabular}
\end{table}

\begin{table}[htbp]
\caption{\label{tab:RK_Th}Determined values of the Kapitza resistance for the 33
investigated (001) GB systems with twist angles $\vartheta$ at $T = 300~\K$. Dimensions of
the simulated nanowires are $L = 47.8~\nm$ and $D = 6.13~\nm$.}
\begin{ruledtabular}
\begin{tabular}{dcdc}
\multicolumn{1}{c}{$\vartheta~(\degr)$} & \multicolumn{1}{c}{$\RK~(\RKu)$} & 
\multicolumn{1}{c}{$\vartheta~(\degr)$} & \multicolumn{1}{c}{$\RK~(\RKu)$} \\ 
\hline
3.75 & $0.34\pm 0.05$ & 48.75 & $1.55\pm 0.02$ \\ 
7.50 & $0.84\pm 0.05$ & 52.50 & $1.53\pm 0.02$ \\ 
11.25 & $0.91\pm 0.05$ & 53.13 & $1.58\pm 0.03$ \\ 
15.00 & $1.12\pm 0.03$ & 56.25 & $1.54\pm 0.02$ \\ 
16.26 & $1.16\pm 0.02$ & 60.00 & $1.55\pm 0.05$ \\ 
18.75 & $1.17\pm 0.03$ & 61.93 & $1.59\pm 0.06$ \\ 
22.50 & $1.29\pm 0.02$ & 63.75 & $1.58\pm 0.05$ \\ 
22.62 & $1.28\pm 0.02$ & 67.38 & $1.52\pm 0.06$ \\ 
26.25 & $1.34\pm 0.05$ & 67.50 & $1.51\pm 0.05$ \\ 
28.07 & $1.42\pm 0.03$ & 71.25 & $1.61\pm 0.05$ \\ 
30.00 & $1.43\pm 0.02$ & 73.74 & $1.55\pm 0.05$ \\ 
33.75 & $1.50\pm 0.02$ & 75.00 & $1.60\pm 0.05$ \\ 
36.87 & $1.49\pm 0.02$ & 78.75 & $1.53\pm 0.05$ \\ 
37.50 & $1.49\pm 0.03$ & 82.50 & $1.52\pm 0.05$ \\ 
41.25 & $1.55\pm 0.02$ & 86.25 & $1.48\pm 0.06$ \\ 
45.00 & $1.55\pm 0.02$ & 90.00 & $1.08\pm 0.04$ \\ 
48.75 & $1.55\pm 0.02$ & 93.75 & $1.48\pm 0.05$ \\ 
\end{tabular}
\end{ruledtabular}
\end{table}

\begin{figure}[htbp]
\begin{center}
  \includegraphics[]{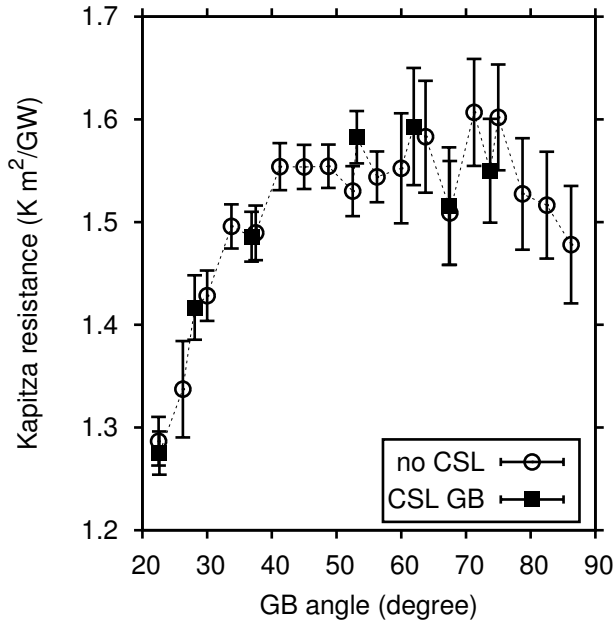}
  \caption{\label{fig:RK_Th_det}Kapitza resistance versus grain boundary angle. Detailed
  excerpt from \Fig{RK_Th} for angles from $22.5\degr$ to $86.25\degr$. The eight
  $\Sigma$ grain boundaries (CSL GB) are marked.}
\end{center}
\end{figure}

Within the error bars, \Fig{RK_Th} shows a monotonous increase of the Kapitza
resistance with twist angle up to $41.25\degr$. For larger angles, i.e., from $41.25\degr$
to $82.5\degr$, $\RK$ is nearly constant with fluctuations about a mean of approximately
$1.55~\RKu$, as can also be seen in more detail in \Fig{RK_Th_det}. We note a drop in
$\RK$ of about $30\%$ for the special $\Sigma 1$ GB at $90\degr$. Besides this case, based
on our results, we can not state with certainty that $\Sigma$ grain boundaries have
distinguished heat conduction properties compared to grain boundaries with similar angles.
While the $51.13\degr$ $\Sigma 5$ GB seems to constitute a local maximum in thermal
boundary resistance, this could easily be caused by statistical uncertainties or
simulation inaccuracies, considering the magnitude of the error bars (see \Fig{RK_Th_det}).
The same applies to the local minima at some of the other $\Sigma$ angles ($36.87\degr$,
$67.38\degr$, and $73.74\degr$). Although it is reasonable to assume modified heat
conduction in CSLs due to the increased periodicity of the superlattice, we can not
confirm this behavior. The effect of enhanced periodicity might be suppressed by the finite 
thickness of the amorphous region and the size of the sectional area not being big enough
 to contain many CSL unit cells.

Schelling \al\cite{Schelling2004} used a method similar to ours, but with periodic
boundary conditions, to simulate (001) Si twist grain boundaries for angles of
$11.42\degr$, and $43.60\degr$. Their results indicate Kapitza resistances of
$0.61~\RKu$, and $1.25~\RKu$, respectively, for $T = 500~\K$. From theory and further
simulations, the thermal boundary resistance is expected to increase for lower
temperatures.\cite{Kimmer2007, Aubry2008} In an estimation, based on results regarding
temperature dependency by Aubry \al\cite{Aubry2008}, we assume that the values of $\RK$
should increase by about $5\%$ to $10\%$, when the temperature decreases from $500~\K$ to
$300~\K$. Our closest angles for comparison are $11.25\degr$, and $45\degr$.
Based on the finite size analysis in \Sub{FS_RK}, we furthermore reduce our $\RK$ values by $17\%$,
which leads to $0.76~\RKu$, and $1.29~\RKu$, respectively. Under the preceding
considerations, our values agree well with those of Schelling \al\cite{Schelling2004},
stated above.

Ju and Liang\cite{Ju2013} studied (001) Si twist grain boundaries by NEMD simulation with
periodic boundary conditions for different temperatures. Their results for $T = 300~\K$
indicate Kapitza resistances of about $1.5~\RKu$, and $2.2~\RKu$ for angles of
$16.26\degr$, and $36.87\degr$, respectively. If we extrapolate our results for these
angles from \Tab{RK_Th} to infinite cross section, our values for $R_\mr{inf}$ reach only
about $60\%$ of Ju and Liang's. The deviations might arise due to different methods used
during the equilibration of the grain boundaries prior to the heat flux phase.
While we use a $NVT$-thermostat for our fixed boundary conditions, they use a
$NPT$-thermostat\cite{Nose1984} (with constant pressure $P$) to relax the system with
periodic boundary conditions. Consequently, in our system, we expect both the pressure to
be higher and a modified structure of the grain boundary in the vicinity of the fixed walls.
However, the differences caused by different configurations should diminish with
increasing cross sectional area. Another reason for the deviation of the results might be
differences in the heat flux generation, as they use the M\"uller-Plathe
method,\cite{Muller-Plathe1997} while we use velocity rescaling. Nevertheless, the values
of $\RK$ are in the same order of magnitude and our simulations yields reasonable
information on the grain boundary angle's influence on thermal boundary resistance.

\section{\label{sec:Concl}Conclusion}

We used NEMD to study thermal transport in crystalline $\Si$ nanowires with and without
(001) twist grain boundaries. A cylindrical geometry was necessary for GB systems with fixed
boundary conditions. Dimensions were in the range of $L \in [24~\nm, 380~\nm]$ and $D \in
[4~\nm, 17~\nm]$. The fixed boundary conditions led to significant finite size effects.
The thermal conductivity increased with both length and diameter, where the former had a much
stronger influence (see Figs.~\fig{kappa_L} and \fig{kappa_D}). The shortest investigated
system yielded a thermal conductivity of about $12~\TCu$, which agrees well with results of
da Cruz \al\cite{daCruz2011}. Increasing this system length eightfold increased $\kappa$
by a factor of $3.5$. The basic configuration for the grain boundary analysis had
dimensions of $L \approx 48~\nm$, and $D \approx 6~\nm$, yielding $\kappa \approx
19~\TCu$. Compared to experimental values for crystalline bulk $\Si$, the thermal conductivity
has been reduced by one order of magnitude\cite{Capinski1997}.

Concerning the thermal Kapitza boundary resistance, we noted very little effect due to length
variations (see \Fig{RK_L}). In contrast, the Kapitza resistance decreased clearly and
systematically with cross sectional area. Our results indicate a linear drop with
inverse cross section (see \Fig{RK_D_inv}). We used this approach to extrapolate to
infinite cross section, for which the influence of the artificial fixed boundary
conditions should vanish. For the $36.87\degr$ $\Sigma 5$ GB with basic dimensions, this
implied an overestimation of $\RK$ by about $17\%$, compared to infinite cross section.
The Kapitza resistance converged quite quickly with diameter, i.e., already for $D \approx
17~\nm$ (approx. $31~a_0$) the deviation from the extrapolation to infinite cross section was reduced to $3\%$. Increasing $D$ further
to $28~\nm$ decreased the deviation to $1\%$. For future simulations with fixed
boundary conditions, we suggest to use diameters of at least this magnitude.

Heat conduction was simulated for 33 different (001) twist grain boundaries from $4\degr$
to $94\degr$ (see \Fig{RK_Th}). Up to $41\degr$, the Kapitza resistance increased with
twist angle. For angles from $41\degr$ to $86\degr$, we could not detect systematic behavior
(see \Fig{RK_Th_det}). The values of $\RK$ remained rather constant, fluctuating between
$1.51~\RKu$ and $1.61~\RKu$. Compared to this range, the thermal boundary resistance dropped
sharply by $30\%$ for the $90\degr$ $\Sigma 1$ GB system. In the context of the calculated
uncertainties, we can neither state nor deny distinguished thermal properties of the other
CSL grain boundaries compared to adjacent angles. It would be desirable to simulate GB
systems with larger diameter for more angles. In doing so, it could be analyzed whether
a special behavior of $\Sigma$ systems is prevented by cross sections which are to small to
contain enough CSL unit cells.

Compared to periodic boundary conditions, we hoped to reproduce more accurately the thermal
processes between individual nano particles with finite dimensions. It would be
enlightening to investigate whether any major shortcomings arise due to the used
artificial fixed boundary conditions. This could best be done by comparing to further
simulations with free boundary conditions. Because our extrapolated values of the Kapitza
resistance agree well with those of Schelling \al,\cite{Schelling2004} we are confident
that our measured values are in a realistic range (cf. \Sub{RK_Th}). At any rate, our
results yield information on the twist angle's qualitative influence on thermal
boundary resistance. Concerning heat conduction in thermoelectric materials composed of
nano particles, we conclude that the relative orientation of the grains should be of
importance. Increasing the number of high angle grain boundaries in polycrystalline
material should reduce thermal conductance of the entire system.

\begin{acknowledgments}
This research has been supported by the German Research Foundation (DFG) (grant no. \mbox{WO577/10-1} and WO577/6-2) within the priority program \mbox{SPP 1386} \textit{Nanostructured Thermoelectric Materials: Theory, Model Systems and Controlled Synthesis}.
\end{acknowledgments}

\bibliography{literature}

\end{document}